\documentstyle[12pt]{article}
\begin{document}
\begin{center}
ANALYTICAL INVESTIGATION OF ANTICIPATING CHAOS SYNCHRONIZATION IN TIME-DELAYED AND CASCADED SYSTEMS \\
E. M. Shahverdiev \footnote{Permanent address: Institute of Physics, 370143 Baku,Azerbaijan},
S.Sivaprakasam and K. A. Shore \footnote{Electronic address: alan@sees.bangor.ac.uk}\\
School of Informatics, University of Wales, Bangor, Dean Street, Bangor, LL57 1UT, Wales, UK\\
\end{center}
For the first time, using a modified Ikeda model it is demonstrated analytically that anticipating synchronization can be obtained in chaotic time-delay systems governed by two characteristic delay times. We derive existence and stability conditions for the dual-time anticipating synchronization manifold. We also show that increased anticipation times for chaotic time-delay systems with two characteristic delay times can be obtained by the use of cascaded systems.\\
~\\
PACS number(s):05.45.Xt, 05.45.Vx, 42.55.Px, 42.65.Sf\\
~\\
\begin{center}
1.Introduction\\
\end{center}
\indent Seminal papers on chaos synchronization [1] have stimulated a wide range of 
research activity in laser physics, electronic circuits, chemical, biological systems, and secure communications; a recent comprehensive review 
of the subject is found in [2]. Time delay systems [3] are ubiquitous in nature and technology, due to finite signal transmission times, switching speeds and memory effects and therefore the 
study of synchronization phenomena in such systems is of great practical importance. 
Time delay systems are interesting because the dimension of their chaotic 
dynamics can be increased by increasing the delay time 
sufficiently [4]. From this point of view these systems are especially 
appealing for secure communication schemes. In addition, time delay 
systems can be considered as a special case of spatio-temporal systems, see e.g. [5] and references therein. Recently [6] it was discovered that dissipative chaotic systems with a 
time-delayed feedback (memory) can drive identical systems in such a way that 
the driven system anticipates the driver by synchronizing with its future 
states. Such a behavior is a result of the interplay between delayed feedback and dissipation [6]. Also it was demonstrated that for small anticipation times, 
anticipating synchronization also occurs in chaotic systems described by ordinary 
differential equations which includes delay due to the finite propagation time of the signal from the driver to the driven system (the so-called coupling delay). Anticipating synchronization [6] appears as a 
coincidence of shifted-in-time 
states of two coupled systems, but in this case, in contrast to lag 
synchronization, the driven system $y$ anticipates the driver $x$,$y(t)=x(t+\tau)$ 
or equivalently $x(t)=y_{\tau}(t)\equiv y(t-\tau$) with $\tau >0$. In [6] anticipating  chaos synchronization was studied in the case of a single delay time. In [7] it is demonstrated that by augmenting the phase space of the driven system (by considering a chain of driven systems), one can accomplish anticipation times that are multiples of the coupling delay time.  Anticipating chaos synchronization for systems with two delay times: a delay in the coupled systems themselves and a coupling delay was investigated {\it numerically} in [8]. The first experimental observation of anticipating synchronization in semiconductor lasers with optical feedback has been reported recently [9]. This experimental work opens up possibilities for practical use of anticipating synchronization phenomenon.
Synchronization of coupled 
chaotic systems restricts the evolution of synchronized systems to the synchronization manifold and therefore eliminates  some degrees of freedom of the joint system, 
thus leading to significant reduction of complexity. In this context from a fundamental point of view, new types of chaos synchronization, including anticipating synchronization can be considered as a novel ways of reducing unpredictability  of chaotic dynamics.
Possible practical applications of anticipating chaos synchronization 
may exploit the fact that driven system {\it anticipates} the driver.For example this 
phenomenon can be used for a fast prediction-because no computation is involved- 
by simply coupling the identical response system to the master system; in secure communications anticipation of the future states of the transmitter (master laser) at the receiver (slave laser) end allows more time to decode the message); 
another possibility can be the control of delay-induced instabilites in a wide range of non-linear systems. Also anticipating synchronization may be of interest for the understanding of natural information processing systems.\\
\indent In this paper, using a modified Ikeda model {\it analytically} generalize the concept of anticipating synchronization to the cases, when there are two delay times in the coupled systems: where the delay time in the coupling is different from the delay time in the coupled systems themselves. We derive existence and stability conditions for the corresponding anticipating synchronization manifold. Furthermore, we show analytically that increased anticipation times for chaotic time-delay systems both with a single delay time in the driver system and with dual delay times can be obtained by the use of cascaded chaotic systems.\\
\begin{center}
2.Anticipating chaos synchronization in time delayed systems with two characteristic delay times\\
\end{center}
For clarity of presentation we reproduce here the definition of anticipating chaos synchronization 
in [6]:\\
The driver system 
$$\hspace*{5cm}\frac{dx}{dt}=-\alpha x + f(x_{\tau}) \hspace*{8.3cm}(1)$$
synchronizes with a driven system of the form
$$\hspace*{5cm}\frac{dy}{dt}=-\alpha y + f(x) \hspace*{8.6cm} (2)$$
on the anticipating synchronization manifold 
$$\hspace*{8cm}x=y_{\tau}.\hspace*{7.5cm}(3)$$
From eqs.(1-2) it follows that 
$\frac{dx}{dt}-\frac{dy_{\tau}}{dt}=-\alpha (x-y_{\tau}) + f(x_{\tau})-f(x_{\tau})=-\alpha (x-y_{\tau})$.
We define the error signal by symbol $\Delta$: $\Delta=x-y_{\tau}$.  Then $\frac{d\Delta}{dt}=-\alpha \Delta$. In many representative cases, chaos 
synchronization can be understood from the existence of a global 
Lyapunov function of the error signals [10]. Thus by introducing the Lyapunov function $L=\frac{1}{2}\Delta^{2}$ we obtain that for $\alpha >0 $ the 
anticipating synchronization manifold $x=y_{\tau}$ is globally attracting and asymptotically stable.\\
Throughout this paper to enhance the accessibility and practicality of our presentation, we confine ourselves to the demonstration of principles using specific examples- the modified Ikeda model [8] and the (conventional) Ikeda model [6].\\
\indent Consider the following modified version of the unidirectionally coupled Ikeda model[8].
$$\hspace*{-5cm}\frac{dx}{dt}=-\alpha x + m_{1} \sin x_{\tau_{1}},$$
$$\hspace*{4cm}\frac{dy}{dt}=-\alpha y + m_{2} \sin y_{\tau_{1}} + m_{3}\sin x_{\tau_{2}},\hspace*{6.5cm}(4)$$
where $\alpha$ is a positive constant; $m_{1},m_{2}$ and $m_{3}$ are constants;
$\tau_{1}$ is the feedback delay in the coupled systems; $\tau_{2}$ is the coupling delay.\\
Now we shall analytically demonstrate that $x=y_{\tau_{1} - \tau_{2}}$ with $\tau_{1} > \tau_{2}$ can be the anticipating synchronization manifold; find the existence and stability conditions for anticipating synchronization, and then compare the analytical results with numerical simulations.\\
From eqs.(4) it follows that under the condition  
$$\hspace*{7cm}m_{1} = m_{2}+ m_{3},\hspace*{6.7cm}(5)$$
the dynamics of the error $\Delta =x-y_{\tau_{1} - \tau_{2}}$ obeys the following equation:
$$\hspace*{6cm}\frac{d\Delta}{dt}=-r\Delta + s \Delta_{\tau_{1}},\hspace*{7.1cm}(6)$$
\noindent with $r=\alpha$ and $s=(m_{1}-m_{3})\cos x_{\tau_{1}}$. 
It is obvious that $\Delta=0$ is the solution of eq.(6). The stability condition for the trivial solution $\Delta=0$ of eq.(6) can be found by investigating the positively defined Krasovskii-Lyapunov fuctional 
$V(t)=\frac{1}{2}\Delta^{2} + \mu\int_{-\tau}^{0}\Delta^{2}(t+t_{1})dt_{1}$ 
(where $\mu >0$ is an arbitrary positive parameter). According to [3-4] the 
sufficient stability condition for the trivial solution of eq.(6) is: $r>\vert s \vert$. Then the sufficient stability condition for the anticipating synchronization manifold $x=y_{\tau_{1} - \tau_{2}}$ reads: 
$$\hspace*{6cm}\alpha > \vert m_{2}\vert.\hspace*{8.8cm}(7)$$
The condition $m_{1}=m_{2} + m_{3}$ is the existence (necessary) condition for anticipating synchronization for the unidirectionally coupled modified Ikeda model.\\
Thus, in this section of the paper, for the first time, we have derived analytically existence and sufficient stability conditions for anticipating synchronization in dual-time coupled modified Ikeda model.\\
\begin{center}
2.Cascaded anticipation of chaos synchronization\\
\end{center}
From the practical application point of view  it is of great importance to achieve larger anticipation times 
between the driver and the driven systems. In this section we demonstrate that cascaded response systems 
configuration can be used to achieve that aim. Synchronization between cascaded systemswas demonstrated experimentally for the first time in [12] in semiconductor laser diodes with optical feedback. In [7] it is demonstrated that by augmenting the phase space of the driven system (by considering a chain of driven systems), one can accomplish anticipation times that are multiples of the coupling delay time.
Here we demonstrate that increased anticipation times for chaotic time-delay systems both with a single delay time in the driver system and with dual delay times also can be obtained by the use of cascaded chaotic systems configuration.\\
Consider the situation when the driven system $y$ in eqs.(1-2) itself drives another 
response system $z$:
$$\hspace*{0.1cm}\frac{dx}{dt}=-\alpha x + f(x_{\tau}),$$
$$\hspace*{0.1cm}\frac{dy}{dt}=-\alpha y + f(x),$$
$$\hspace*{7cm}\frac{dz}{dt}=-\alpha z + f(y).\hspace*{6.4cm} (8)$$
We demonstrate analytically that the driven system $z$ synchronizes with the driver system 
$x$ with the anticipation time $2\tau$. Let us calculate the following difference:
$\frac{dx}{dt}-\frac{dz_{2\tau}}{dt}=-\alpha (x-z_{2\tau})+ f(x_{\tau})-f(y_{2\tau})$.
Assume that  anticipiating synchronization between $x$ and $y$ state variables 
has already taken place; then from $x=y_{\tau}$ we obtain that $x_{\tau}=y_{2\tau}$.
Then we arrive at the error $\Delta=x-z_{2\tau}$  dynamics  $\frac{d\Delta}{dt}=-\alpha \Delta$. In other words having two driven systems it is possible to double the anticipation time. It is straightforward to verify that having $n$ driven systems allows for the anticipation times $n\tau$. Thus, using cascaded driven systems it is possible to obtain the anticipation times that are multiples of the delay time in the coupled systems themselves; As mentioned above the anticipation times that are multiples of the coupling delay time is accomplished in [7]. We consider cascaded anticipating synchronization in the following coupled Ikeda systems with a single delay [6]:
$$\hspace*{0.7cm}\frac{dx}{dt}=-\alpha x - \beta \sin x_{\tau},$$
$$\hspace*{0.6cm}\frac{dy}{dt}=-\alpha y - \beta \sin x,$$
$$\hspace*{7cm}\frac{dz}{dt}=-\alpha z - \beta \sin y,\hspace*{6cm}(9)$$
where $\alpha >0, \beta >0$. Using the error dynamics approach one can find that $x=y_{\tau}$ and $x=z_{2\tau}$ are the anticipating chaos synchronization manifolds for the system (9).\\
Next we investigate the possibility of obtaining large anticipating times for chaotic systems with 
{\it dual} delay times. Let the driven system $y$ from eqs.(4) drive another response system $z$:
$$\hspace*{5cm}\frac{dz}{dt}=-\alpha z + m_{4} \sin z_{\tau_{1}} + m_{5}\sin y_{\tau_{2}}.\hspace*{5.5cm}(10)$$
We shall show that systems (4) and (10) provide an anticipation time of 
$2(\tau_{1} - \tau_{2})$.
To verify this, we investigate the error dynamics for $\Delta=x-z_{2(\tau_{1}-\tau_{2})}$:
$$\hspace*{5cm}\frac{d\Delta}{dt}=-\alpha \Delta + m_{1} \sin x_{\tau_{1}} - m_{4}\sin z_{3\tau_{1}-2\tau_{2}} - m_{5} \sin y_{2\tau_{1}-\tau_{2}}.\hspace*{2cm}(11)$$ 
Assuming that synchronization between the driver $x$ and slave system $y$ already had taken place, i.e. using $x=y_{\tau_{1} - \tau_{2}}$ and $y_{2\tau_{1} - \tau_{2}}=x_{\tau_{1}}$, under the existence condition $m_{1}=m_{4} + m_{5}$ we obtain also a sufficient stability 
condition for the anticipating synchronization manifold $x=z_{2(\tau_{1}-\tau_{2})}$,
 $\alpha > \vert m_{4}\vert$. It is clear that, having $n$ response systems one can obtain the anticipation time $n(\tau_{1} - \tau_{2})$.\\
Thus, we have demonstrated that increased number of driven systems will allow for larger anticipitation times for the coupled dual time chaotic systems.\\ 
\indent In conclusion, we have analytically investigated the phenomenon of anticipating 
synchronization in unidirectionally coupled time delayed modified Ikeda systems with two characteristic delay times. We have obtained that the anticipation time is the 
difference between the delay time in the coupled systems and the coupling delay time and derived both 
existence and sufficient stability conditions for the anticipating synchronization manifold.
In order to exploit the capability of {\it anticipating future states} of the master system, it is of great importance to obtain increased anticipating times. We have demonstrated here that the concept of cascaded slave systems can provide large anticipating times for the dual time chaotic time delay systems.\\
This work is supported by UK EPSRC under grants GR/R22568/01 and GR/N63093/01.\\

\end{document}